\magnification=1200
\voffset=.5truecm
\hoffset=-.5truecm
\vsize=24truecm
\hsize=17truecm
\parskip=4pt
\null

\font\tbf=cmbx10 scaled \magstep2
\hyphenation {Schwarz-schild}
\hyphenation {Abra-mo-wicz}
\font\tbmi=cmmib10 
\def\blambda{\hbox{\tbmi\char'025}}
 
\def\btimes{\hbox{\tbs\char'002}}
\def\bnabla{\hbox{\tbs\char'162}}
\font\tenmib=cmmib10 \textfont"E=\tenmib
\font\tenbsy=cmbsy10 \textfont"F=\tenbsy

\mathchardef\blambda="0E15
\mathchardef\bnabla="0F72
\mathchardef\btimes="2F02

\def\ref{\par\noindent\hangindent=1 truecm}
\def\simless{\mathbin{\lower 3pt\hbox
   {$\rlap{\raise 5pt\hbox{$\char'074$}}\mathchar"7218$}}} 
\def\simgreat{\mathbin{\lower 3pt\hbox
   {$\rlap{\raise 5pt\hbox{$\char'076$}}\mathchar"7218$}}} 
\null
\noindent
{\tbf Rayleigh and Solberg criteria reversal near black holes:
the optical geometry explanation}
\vskip 0.8truecm
\par \noindent
{\tbf Marek A. Abramowicz}
\vskip 0.2 truecm \par \noindent
\ref {\it Nordita, Copenhagen, Denmark\footnote {$^*$}{\sevenrm
On leave from Department od Astrophysics, G\"oteborg University. Address: Theoretical Physics, Chalmers University, 412-96 G\"oteborg, Sweden}}

\vskip 0.7truecm
\par \noindent 
\vskip 0.8truecm
\noindent \hskip 2,5truecm
{\bf Summary}
\par \noindent \hskip 2.5truecm
{\hangindent 2.5truecm 
{The familiar Newtonian version of the Rayleigh criterion demands that
for dynamical stability the specific angular momentum should increase
with the increasing circumferential radius of circular trajectories of
matter. However, sufficiently close to a black hole the Rayleigh
criterion reverses: in stable fluids, the angular momentum must
decrease with the increasing circumferential radius. The geometrical
reason for this reversal is that the space is so very strongly warped
that it turns inside-out. \par}}
\par 
\vskip 0.2truecm
\noindent \hskip 2,5truecm
{\hangindent 2.5truecm {\bf Key words:} hydrodynamics: instabilities --
relativity -- optical reference geometry \par}
\vskip 2truecm

\noindent
{\bf 1 INTRODUCTION}
\vskip 0.5truecm

\noindent
Abramowicz and Prasanna (1990) found that, surprisingly, in the
region\footnote {$^*$}{\sevenrm In this paper I use the c $= 1 =$ G
units and the $+~-~-~-$ signature.} $r < 3M$ around of a nonrotating
(Schwarzschild) black hole with the mass $M$, the necessary condition
for stability of axially symmetric, rotating isentropic and
incompressible fluid against small, adiabatic and axially symmetric
perturbations is that the specific (per unit mass) angular momentum of
the fluid $j$ should {\it decrease} with the increasing circumferential
radius of the circular trajectories of the fluid $r$,
$$ \left ( {{\partial j^2}\over {\partial r}}\right ) < 0.
\eqno (1.1)$$
The result was very surprising indeed, because it seemingly
contradicted the old classical result of Rayleigh that, in Newton's
theory, the necessary condition for stability should be,
$$ \left ( {{\partial j^2}\over {\partial r}}\right ) > 0.
\eqno (1.2)$$
Already a long time ago the Rayleigh criterion (1.2) was generalized to
isentropic, compressible fluids by Solberg, and to non-isentropic
fluids by H{\o}iland. Rayleigh and Solberg criteria, which both have
identical form (1.2), can be derived from the more general H{\o}iland
criterion as special cases (see Tassoul, 1978 for a detailed
description of these criteria). Relativistic version of the H{\o}iland
criterion has been derived with a great care, and in a very
mathematically elegant way, in an important paper by Seguin (1975), but
the effect of the Rayleigh and Solberg criteria reversal has not been
noticed there.

\noindent
The reversal of Rayleigh and Solberg stability criteria provides yet
another nontrivial example of an ambiguity of the geometrical meaning
of ``inside'' and ``outside'' directions. I pointed out a few years ago
(Abramowicz, 1992) that several counter intuitive effects in dynamics
of bodies closely orbiting black holes could be understood as resulting
from a strong warping of geometry that literally turns the space inside
out! In the present paper, I show that the surprising effect of the
Rayleigh and Solberg criteria reversal may also be explained the same
way. 

\vskip0.5truecm
\noindent
{\bf 2 THE LICHTENSTEIN THEOREM}
\vskip 0.3truecm

\noindent
My discussion will be restricted to the motion on the equatorial
symmetry plane of the rotating fluid. This restriction is {\it not}
relevant for the results, but it greatly simplifies some arguments.

\noindent
Lichtenstein (1933) has proved that in Newton's theory
stationary rotating fluid bodies always display a mirror symmetry with
respect to reflections in a plane orthogonal to the rotating axis and
containing the barycenter of the body (the equatorial plane). For
stationary, axially symmetric fluids, the equatorial plane is itself
rotationally symmetric, and thus one may introduce on it a family of
concentric circles around the symmetry centre. The circumferential
radii $r$ of these circles provide a convenient radial coordinate on the
equatorial plane.

\noindent
In the early 1970s I was a Ph.D.
student of Andrzej Trautman in Warsaw. He used to give his students a
lot of freedom in performing research, and several of us have been
indeed involved in doing all kind of esoteric theoretical physics
in addition to our official Ph.D.
subjects. Chandrasekhar who was often coming to visit Trautman in
Warsaw, asked me once what I was working on in addition to my main
subject. I answered, proudly, that I was trying to prove Lichtenstein's
theorem in general relativity.  Chandra could not hide his obviously
disapproving surprise, and only after a moment of silence replied very
politely, ``This sounds a bit too formal to me''. Of course, when
Chandra left, Trautman told me to stop with Lichtenstein and do
something less formal.

\noindent
It is still unknown whether the Lichtenstein theorem is valid in
Einstein's relativity,
but all the known solutions of Einstein's field
equations describing stationary, axially symmetric spacetimes do show
this discrete symmetry. For example, the Kerr solution in
Boyer-Lindquist coordinates is obviously symmetric with respect to
reflection $\theta \rightarrow \pi - \theta$ with respect to the
equatorial plane, $\theta = \pi/2$,

\vfill \eject \null

\vskip0.5truecm
\noindent
{\bf 3 RAYLEIGH AND SOLBERG CRITERIA IN NEWTON'S THEORY}
\vskip 0.3truecm

\noindent
In this Section I remind the familiar heuristic proof of the Rayleigh
and Solberg criteria, first given by Randers (1942), and in the next
Section I use a slightly modified proof to derive the relativistic
version of the criteria. 

\noindent
In a reference frame which corotates with the small fluid element under
consideration, three forces are acting on the element: gravitational
force ${G}$, centrifugal force ${Z}$, and applied force (e.g.
pressure) ${T}$. On a particular circle $r = r_0$ it is,
$$ {G}(r_0) + {Z}(r_0, j^2_0) + {T}(r_0) = 0.  \eqno (3.1)$$

\noindent
Consider an adiabatic, axially symmetric perturbation in which the
small element initially located at $r_0$ is displaced to $r = r_0 +
\delta r$. As the perturbation is adiabatic and axially symmetric, it
does not change the specific angular momentum of the element. The
(squared) angular momentum at the perturbed position $r = r_0 + \delta
r$ is therefore $j_0^2$. On the other hand, the (squared) specific
angular momentum needed at this very place for the balance of the
forces (3.1) should be,
$$ j^2 = j^2 (r_0 + \delta r) = j^2_0 + \delta j^2 = j^2_0 + \left (
{{\partial j^2}\over {\partial r}}\right ) \delta r. \eqno (3.2)$$
Because $j_0^2 \not = j^2$, there is unbalanced force $\delta T$ at
$r_0 + \delta r$,
$$ T(r)  + G(r) + Z(r, j^2 - \delta j^2) =
\delta T. \eqno (3.3) $$
Subtracting (3.1) from (3.3) and using (3.2) one gets,
$$ \delta T = - \left ( {{\partial Z}\over {\partial j^2}}\right )
\left ( {{\partial j^2}\over {\partial r}}\right ) \delta r. \eqno
(3.4)$$
Of course, the perturbation could be stable {\it only if} the
unbalanced force brings the perturbed element back to its original
position, {\it i.e.} only if $\delta T \delta r < 0$. Thus, the
necessary condition for stability yields,
$$  \left ( {{\partial Z}\over {\partial j^2}}\right )
\left ( {{\partial j^2}\over {\partial r}}\right ) >
0.\eqno (3.5)$$

\noindent
In the corotating frame, according to Newton's theory, $Z(r, j^2) =
j^2/r^3$. Therefore, in Newton's theory, ${{\partial Z}/{\partial
j^2}}$ is unconditionally positive,
$$ \left ( {{\partial Z}\over {\partial j^2}}\right ) = {1\over r^3} >
0, \eqno (3.6)$$
and this is why the Rayleigh and Solberg criteria take the familiar
form (1.2). However, Abramowicz and Prasanna (1990) have shown that,
according to Einstein's theory of general relativity, it could be that
${{\partial Z}/{\partial j^2}} < 0$. In this case the Rayleigh and
Solberg criteria demand that for stability $\partial j^2/\partial r <
0$.

\vskip0.5truecm
\noindent
{\bf 4 RAYLEIGH AND SOLBERG CRITERIA IN EINSTEIN'S THEORY}
\vskip 0.3truecm

\noindent
The Rayleigh, Solberg and H{\o}iland criteria are valid for stationary
and axially symmetric equilibria of rotating fluids. Thus, their
generalization to Einstein's theory should be derived in stationary and
axially symmetric spacetimes. In such spacetimes two commuting Killing
vector fields exist: asymptotically timelike Killing vector $\eta^i$
(corresponding to stationarity), and spacelike vector $\xi^i$ with
closed trajectories (corresponding to axial symmetry).  The {\it
stationary observers} are defined as those with velocities $N^i = 
\eta^i(\eta \eta)^{-1/2}$. More convenient are {\it locally non
rotating observers} with four velocities,
$$ n^i = {\rm e}^{\Phi}(\eta^i + \omega \xi^i), \eqno (4.1)$$
$$ {\rm e}^{-2\Phi} = (\eta \eta) - \omega (\xi \xi),~~~
\omega = - {{(\eta\eta)}\over {(\xi\xi)}}. \eqno (4.2)$$
One may check that $n^i$ is a unit timelike vector which has zero
vorticity and is hypersurfaces orthogonal. The hypersurfaces orthogonal
to trajectories of $n^i$ are three dimensional spaces $t={\rm const}$
with the metric,
$$ h_{ik} = g_{ik} - n_i n_k.  \eqno (4.3)$$
Because $(n \xi) = 0$, the three dimensional space is axially symmetric,
and a circle in space can be defined as a trajectory of the projection
of $\xi^i$ into space. A particle which moves along a circle in {\it
space} has its four velocity in the {\it spacetime} given by,
$$ u^i = \gamma (n^i + v \tau^i). \eqno (4.4)$$
Here $\tau^i = \xi^i/r$, with $r^2= -(\xi \xi)$ is a unit vector
orthogonal to $n^i$, and $v$ is the orbital speed.  From
$(uu)=(nn)=1,~(\tau\tau)=-1$ one deduces that $\gamma = 1/(1 -
v^2)^{1/2}$, so that $\gamma$ is the Lorentz redshift factor. Note that
the projection of the four velocity $u^i$ into the 3-D space gives,
${\tilde v}^k = u^i h^k_{~i} = \gamma v \tau^k$ and therefore ${\tilde
v}^2 = - ({\tilde v}{\tilde v}) = \gamma^2 v^2$. While the velocity $v$
changes between $\pm 1$, the ``velocity'' ${\tilde v}$ changes between
$\pm \infty$.

\noindent
Abramowicz, Nurowski and Wex (1993) demonstrated that in the corotating
reference frame of matter that moves according to (4.4), the
acceleration $a_i = u^k \nabla_k u_i$  equals,
$$ a_i = \nabla_i \Phi + \gamma^2{v^2\over R}\nabla_i R - \gamma^2 v
R\nabla_i \omega, \eqno (4.5)$$
where $R$ is the radius of gyration $R$ defined as (Abramowicz,
Miller and Stuchl{\'\i}k, 1993),
$$ R = r{\rm e}^{\Phi} = [-(\xi\xi)]^{1/2}{\rm e}^{\Phi}.
\eqno (4.6)$$

\noindent
Abramowicz, Nurowski and Wex explained that the acceleration formula
(4.5) suggests a very natural and convenient way to covariantly define
gravitational, centrifugal and Lense-Thirring (``Coriollis'') forces,
$$ {\cal G}_i = \nabla_i \Phi,~~{\cal Z}_i = \gamma^2{v^2\over
R}\nabla_i R,~~ {\cal C}_i = - \gamma^2 v R\nabla_i \omega. \eqno
(4.7)$$

\noindent
In order to repeat dervivation of the stability criterion in a way
similat to that discussed in the previous Section, one needs to know
how to express the orbital speed by angular momentum which is conserved
during an axially symmetric, non-stationary perturbation.  In
relativity, exactly as in Newton's theory, orbital speed $v$, and
specific angular momentum per unit energy $\ell$,
$$\ell = - {{(u\xi)}\over {(u\eta)}}, \eqno (4.8)$$
are connected by  the ``gyration equation'',
$$ v = {{\ell}\over R}. \eqno (4.9)$$
The specific angular momentum per unit energy $\ell$ is {\it not}, in
general, conserved by axially symmetric perturbations, because such
perturbations need not to be stationary. The quantity conserved in
axially symmetric perturbations (including nonstationary ones) is the
specific angular momentum per unit mass,
$$ j = \ell {\cal E}, \eqno (4.10)$$
where ${\cal E}$ is the specific {\it per unit mass} energy. In
Newton's theory $j = \ell$ for any fluid, but in relativity $j \not =
\ell$. However, in the special case of isentropic fluid considered
here, the relativistic von Zeipel theorem (Abramowicz, 1974) guarantees
that ${\cal E}={\cal E}(\ell),~ j=j(\ell)$, and
$$ {{dj^2}\over{d\ell^2}}= {{\cal E}^2\over {1-\Omega \ell}} > 0. \eqno
(4.11)$$
Here $\Omega$ is the orbital angular velocity measured by stationary
observers. It is connected to the orbital angular velocity measured by
locally non rotating observers ${\tilde \Omega}$ by, ${\tilde \Omega}=
\Omega - \omega$, where
$$ {\tilde \Omega} = {v^2\over \ell} = {v \over R} = {\ell \over R^2}.
\eqno (4.12)$$

\noindent
Using general expressions (4.7) for inertial forces, gyration equations
(4.9), and assuming that the fluid is isentropic, I write on the
equatorial plane,
$$ {\cal G}(r) = {{\partial \Phi}\over {\partial r}}, \eqno (4.13)$$
$$ {\cal Z}\left(r, \ell^2(j)\right) = {{\ell^2(j)}\over {R^2 -
\ell^2(j)}}{1\over R^3} {{\partial R}\over {\partial r}}, \eqno (4.14)$$
$$ {\cal C}\left(r, \ell(j)\right) = - {{\ell(j) R^2}\over {R^2 -
\ell^2(j)}} {{\partial \omega}\over {\partial r}}. \eqno (4.15)$$
It should be obvious that the acceleration formula (4.9) may now be
written on the equatorial plane in the form similar to the Newtonian
equation (3.1),
$$ {\cal G}(r_0) + {\cal Z}\left(r_0, \ell^2(j_0)\right) 
+ {\cal C}\left(r_0, \ell(j_0)\right) + {\cal T}(r_0) = 0.  \eqno (4.16)$$
The additional Lense-Thirring force ${\cal C}$ is not present in
Newtonian case, and one must remember that the quantity conserved in
the perturbation is not $\ell_0$ but $j_0$. Taking into account that
${{\partial j^2}/{\partial \ell^2}}>0$ according to (4.11), one writes
the general form of the criterion in the form,
$$ \left[ \left ( {{\partial {\cal Z}}\over {\partial \ell^2}}\right )
+ \left ( {1\over {2\ell}}\right )
\left ( {{\partial {\cal C}}\over {\partial \ell}}\right )\right ] 
\left ( {{\partial \ell^2}\over {\partial r}}\right ) >
0.\eqno (4.17)$$
After a tedious calculation that utilises gyration equation one can
rearrange this and write,
$$ \left( {{\partial \ln R}\over{\partial r}} - {{1 + v^2}\over
{2{\tilde \Omega}}}{{\partial \omega}\over {\partial r}}
\right) \left({{\partial \ell^2}\over {\partial r}}\right) > 0. \eqno (4.18)$$
Seguin (1975) wrote the criterion in a different, but equivalent form,
which follows from (4.17) also after tedious calculations. In the
notation adopted here it yields,
$$ \left( {{\partial \ell}\over{\partial r}} - R^2 {{\partial
\Omega}\over {\partial r}}
\right) \left({{\partial \ell}\over {\partial r}}
\right) > 0. \eqno (4.19)$$
Although (4.18) and (4.19) are fully equivalent, it is convenient to
discuss physical and geometrical meaning of the criterion using (4.18).  

\vskip0.5truecm
\noindent
{\bf 5 REVERSAL OF THE RAYLEIGH AND SOLBERG CRITERIA}
\vskip 0.3truecm

\noindent
The reversal of the Rayleigh and Solberg criteria occurs when the
quantity 
$$ {\cal Q}(r, v) \equiv
v{{\partial R}\over{\partial r}} - 
{1\over2}(1 + v^2)R^2{{\partial \omega}\over {\partial r}} \eqno (5.1)$$
changes its sign, {\it i.e.} at the circle ${\cal Q}(r, v)=0$. I shall
explain the physical and geometrical meaning of ${\cal Q}=0$ by
considering precession of gyroscopes moving on circular orbits, and
circular free motion of photons.

\noindent
The spin of an orbiting gyroscope does precess. The precession rate
with respect to the vector $\tau^i$ tangent to the circle is given by
the Fermi-Walker derivative, $\omega^*_i = u^j\nabla_j u_i - (u^j a_i -
u_i a^j)\tau_j$.  On the equatorial plane, this vector has only one
nonzero component which equals (Abramowicz, 1993),
$$ \omega^* = \gamma^3 \left[ -{\tilde \Omega}{{\partial R}\over
{\partial r}} + {1\over 2}(1 +v^2)R{{\partial \omega}\over {\partial
r}} \right]. \eqno (5.2)$$
The factor $\gamma^3$ in the front of the bracket is due to special
relativistic effects and represents the well-known Thomas precession.
The first term in the bracket represents the geodesic (de Sitter)
precession and the second term is the gravomagnetic (Lense-Thirring)
precession\footnote{$^{\ddagger}$}{\sevenrm There is a considerable
confusion in the literature about how one should interpret precession
of an orbiting gyroscope in terms of different kinds of simple
precessions. In particular, some authors have previously been
discussing a rather unfortunate concept of the ``gravitational Thomas
precession''. This was because in approximate formulae used previously by
several authors, the simple and unique division
provided by (5.2) was not apparent.}. In a weak
gravitational field, ${\partial R}/{\partial r} \approx 1 > 0$, and
${\partial R}/{\partial r} \gg \vert {\partial \omega}/{\partial
r}\vert$, so that $\omega^*$ and ${\tilde \Omega}$ have opposite signs.
This is exactly what is expected by intuition which tells that an
orbiting gyroscope must precess {\it backward} with respect to its
orbital motion in order to always point to the same direction in space.
Thus, gyroscopes orbiting clockwise precess anti-clockwise and vice
versa. However, surprisingly, the bracket in (5.2) may change its sign.
When this happens, the sense of the precession is exactly opposite to
what is expected: gyroscopes are precessing {\it forward} --- those on
clockwise orbits are precessing clockwise, and those on anti-clockwise
orbits are precessing anti-clockwise. (Rindler and Perlick, 1990;
Abramowicz 1990, 1993; Nayak and Vishveshwara 1998). It is not difficult
to
see that the reversal of the Rayleigh criterion and the reversal of the
sense of gyroscope precession go hand in glove. Indeed, (5.2) may be
written as,
$$ \omega^* = - \gamma^3 {1\over R} {\cal Q}, \eqno (5.3)$$
which means that both reversals are governed by exactly the same
condition ${\cal Q}=0$.

\noindent
To see what does this condition really mean, one should consider the
free (geodesic) motion of photons which is described by general
equations $a_i = 0$, and $v = \pm 1$. It follows directly from (4.9),
that they are equivalent to
$$ {{\partial R}\over{\partial r}} \pm R^2{{\partial \omega}\over
{\partial r}} = 0. \eqno (5.3)$$
One should note that (5.3) is the same as the ultra-relativistic limit,
$v = \pm 1$, of both (5.1) and (5.2). Thus, for the ultra-relativistic
fluid, both reversals~--- of the Rayleigh and Solberg criteria and of
the sense of the precession of gyroscopes~--- occur at the location of
the circular photon orbit. The conclusion that the reversal of the
sense of gyroscope precession occurs at the location of a circular
photon orbit has been previously reached by Nayak and Vishveshwara, 1998.

\vskip 0.5truecm
\noindent
{\bf 6 DISCUSSION AND CONCLUSIONS}
\vskip 0.3truecm

\noindent
Abramowicz, Carter and Lasota (1988) have noticed that dynamics of
particles and photons looks simple when described in a conformally
rescaled three dimensional geometry of space with the metric,
$$  {\tilde h}_{ik} = {\rm e}^{-2\Phi}h_{ik}. \eqno (6.1)$$
The conformally rescaled metric (6.1) is called the optical reference
geometry\footnote {$^{\S}$}{\sevenrm The optical reference geometry
(6.1) has been introduced by Dowker and Kennedy (1978) and further
discussed by Gibbons and Perry (1978), and Kennedy, Critchely, and
Dowker (1980). It should be not confused with the optical geometry
introduced by Trautman (1984) and Robinson and Trautman (1984) which is
a totally different concept.}. 

\noindent
Consider a static spacetime, $\omega = 0$. Its metric can be written in
the form,
$$ ds^2 = {\rm e}^{-2\Phi}\left( dt^2 + {\tilde h}_{ik}dx^i dx^k\right),
\eqno (6.2)$$
where the time coordinate $t$ is invariantly defined as the synchronized time
of the static observers, who have four velocities given by $N^i = n^i$. It
is, $\delta^i_{\; (t)} = \eta^i$. Relativistic Fermat's principle
assures that $\int dt$ on the light trajectory has an extremal value,
$\delta \int dt= 0$.  This, together with $dt^2 = - {\tilde h}_{ik}dx^i
dx^k$ that follows from (6.2) and the fact that $ds=0$ for light, is
equivalent to say that for any light trajectory $x^i=x^i(p)$ in the
space of optical reference geometry,
$$ \delta \int \left( - {\tilde h}_{ik}{dx^i\over dp} {dx^k\over
dp}\right)^{1\over2}dp = 0. \eqno (6.3)$$
This means that light trajectories are {\it geodesic lines} in the
optical reference geometry, and this is the reason for its name.

\noindent
The circumference radius of a $r = {\rm const}$ circle equals, in the
optical reference geometry, to the radius of gyration of the circle,
because both radii are given by the same formula, $R=r{\rm e}^{\Phi}$.
I shall now calculate the radius of the curvature of the circle in
optical geometry, ${\tilde {\cal R}}$, using the well-known Frenet
construction (see {\it e.g.} Synge and Schild, 1959),
$$ {\tilde \tau}^i {\tilde \nabla}_i {\tilde \tau}_k = - {\tilde
\lambda}_k {1\over {\tilde {\cal R}}}. \eqno (6.4)$$
Here ${\tilde \lambda}$ is the unit, outside pointing, vector orthogonal
to the circle, ${\tilde \tau}^i$ is the unit vector tangent to the
circle,
$$ {\tilde \tau}^i = {\rm e}^{-\Phi},~~~{\tilde \tau}_i = {\rm
e}^{\Phi},
\eqno (6.5)$$
and the covariant derivative operator ${\tilde \nabla}_i$ obeys,
$$ {\tilde \tau}^i {\tilde \nabla}_i {\tilde \tau}_k =
{\tau}^i {\nabla}_i {\tau}_k + \nabla_k \Phi + \tau_k \tau^i\nabla_i\Phi.
\eqno (6.6)$$
From (6.4) I derive,
$${{\partial R}\over{\partial r}} = \varepsilon {R \over{\cal R}}, 
\eqno (6.7)$$
where $\varepsilon = {\rm sign} ({\partial R}/{\partial r})$.  The formula
for the gyroscope precession takes now the form,
$$ \omega^* =  \gamma^3\left[ -\varepsilon {R\over {\cal R}}{\tilde
\Omega} + {1\over2}(1 + v^2)R{{\partial \omega}\over {\partial
r}}\right ].
\eqno (6.8)$$
In a curved space, consider a circle with circumferential radius $R$
and curvature radius ${{\cal R}}$. If $R\not = {\cal R}$, a simple
geometrical construction reveals that there is a
nonzero deficit angle,
$$ \delta \phi = 2\pi {{{\cal R} - R}\over {\cal R}}. \eqno (6.9)$$
In the case of a very slow orbital motion, $\gamma = 1$, in the static
space, $\omega(r) = 0$, the precession of a gyroscope orbiting the
circe with a steady angular velocity $\Omega = 2\pi/T$ ($T$ is the
orbital period) is obviously given by,
$$ \omega^* =  {{2\pi - \delta \phi}\over T} = -\Omega {R\over {\cal
R}}. \eqno (6.10)$$
This is exactly the first term in formula (6.8). Thus, this term is
{\it indeed} just exactly the geodesic precession~--- but ``geodesic''
in the sense of the optical reference geometry ${\tilde h}_{ik}$, not
the directly projected geometry $h_{ik}$.

\noindent
In this article, I have described a rather counter intuitive effect of
the reversal of the Rayleigh and Solberg stability criteria, and
explained that it could be understood in terms of a very simple
intuitive argument with a help of the optical reference geometry. In
addition to this effect, the optical reference geometry has already
successfully explained a score of similarly strange other effects. One
should ask therefore the question: {\it why} is the optical reference
geometry, and not the ``real'' geometry, so excidingley well fitted to
behaviour of particles, fluids, charges, photons, gravitational waves,
electromagnetic fields and
gyroscopes? Are we
missing something? Is the optical geometry just a particularly good map
of the real curved geometry, or is the optical geometry itself the real
geometry of space?

\vskip0.5truecm
\noindent
{\bf ACKNOWLEDGEMENTS}
\vskip 0.3truecm

\noindent
This research was supported by Nordita througth the Nordic Project,
{\it Non-linear Phenomena in Accretion Discs Orbiting Black Holes}. I
would like to thank Sebastiano Sonego for many very helpful
suggestions.

\vskip0.5truecm
\noindent
{\bf REFERENCES}
\vskip 0.3truecm

\ref Abramowicz M.A., 1974, {\it Acta Astr.}, {\bf 24}, 45
\ref Abramowicz M.A., 1990, {\it Mon. Not. R. astr. Soc.}, {\bf 245}, 733
\ref Abramowicz M.A., 1992, {\it Mon. Not. R. astr. Soc.}, {\bf 256}, 710
\ref Abramowicz M.A., 1993, in {\it The Renaissance of General Relativity and
     Cosmology}, eds. G. Ellis, A. Lanza and J. Miller, Cambridge
     University Press, (Cambridge)
\ref Abramowicz M.A., Carter B. and Lasota J.-P., 1988, {\it Gen. Relat.
     Grav.}, {\bf 20}, 1173
\ref Abramowicz M.A., Miller J.C. and Stuchl{\'\i}k Z., 1993, 
     {\it Phys. Rev. D.}, {\bf 47}, 1440
\ref Abramowicz M.A., Nurowski P., and Wex N., 1993, {\it Class. 
     Quantum Grav.}, {\bf 10}, L183
\ref Abramowicz M.A. and Prasanna A.R., 1989, {\it Mon. Not. R. astr. Soc.},
     {\bf 245}, 270 
\ref Dowker J.S., and Kennedy G., 1978, {\it J. Phys. A}, {\bf 11}, 895
\ref Gibbons G.W., and Perry J.M., 1978, {\it Proc. Roy. Soc. Lond. A},
     {\bf 358}, 467
\ref Kennedy G., Critchely R., Dowker J.S., 1980, {\it Ann. Phys. NY},
     {\bf 125}, 346        
\ref Lichtenstein L., 1933, {\it Gleich\-gewichts\-figuren Rotierender
     Fl{\"u}ssid\-keiten}, Springer Verlag, (Ber\-lin) 
\ref Nayak K.R., and Vishveshwara C.V., 1998, {\it Gen. Rel. Grav.},
     {\bf 30}, 593
\ref Randers G., 1942, {\it Astrophys. J.}, {\bf 95}, 454
\ref Rindler W., and Perlick C., 1990, {\it Gen. Rel. Grav.}, {\bf 22}, 1067
\ref Robinson I. and Trautman A., 1986, in {\it Les Th{\'e}ories de la
     Gravitation}, {\'E}ditions du CNRS (Paris) 
\ref Seguin F.H., 1975, {\it Astrophys. J.}, {\bf 197}, 745
\ref Semerak O., 1997, {\it Gen. Rel. Grav.}, {\bf 13}, 2987
\ref Synge J.L. and Schild A., 1959, {\it Tensor Calculus}, University of
     Toronto Press, (Toronto) 
\ref Tassoul J.-L., 1978, {\it Theory of Rotating Stars}, Princeton
     University Press (Princeton)
\ref Trautman A., 1984, {\it J. Geom. Phys.}, {\bf 1}, 85

\bye